\documentclass[apj]{emulateapj}

\defcitealias{bloom98}      {1}
\defcitealias{djorgovski01} {2}
\defcitealias{djorgovski03} {3}
\defcitealias{djorgovski98} {4}
\defcitealias{christensen05}{5}
\defcitealias{le floc'h02}  {6}
\defcitealias{chary02}      {7}

\shorttitle{Star formation rates and stellar masses in $z$ $\sim$ 1 GRB hosts}
\shortauthors{Castro Cer\'on et al.}
\slugcomment{The Astrophysical Journal, 653: L85--L88, 2006 December 20}

\begin{document}

\title{Star formation rates and stellar masses in $z$ $\sim$ 1 gamma ray burst hosts\altaffilmark{1}}

\author{J.M. Castro Cer\'on\altaffilmark{2},
  M.J. Micha{\l}owski\altaffilmark{2},
          J. Hjorth\altaffilmark{2},
          D. Watson\altaffilmark{2},
      J.P.U. Fynbo\altaffilmark{2}
      and J. Gorosabel\altaffilmark{3}
       }

\altaffiltext{1}{This work is based in part on observations made with the {\em Spitzer Space Telescope}, which is operated by the Jet Propulsion Laboratory, California Institute of Technology, under a contract with NASA.}
\altaffiltext{2}{Dark Cosmology Centre, Niels Bohr Institute, University of Copenhagen, Juliane Maries Vej 30, DK-2100 Copenhagen \O, Denmark; josemari@dark-cosmology.dk, michal@dark-cosmology.dk, jens@dark-cosmology.dk, darach@dark-cosmology.dk, jfynbo@dark-cosmology.dk.}
\altaffiltext{3}{Instituto de Astrof\'{\i}sica de Andaluc\'{\i}a (CSIC), c/. Camino Bajo de Hu\'etor, 50, E-18.008 Granada, Spain; jgu@iaa.es.}

\begin{abstract}

We analyse 4.5, 8 and 24\,$\mu$m band $Spitzer$ images of six gamma ray burst host galaxies at redshifts close to 1. We constrain their star formation rates (SFR) based on the entire available spectral energy distribution rather than the 24\,$\mu$m band only. Further, we estimate their stellar masses ($M_\star$) based on rest frame $K$ band luminosities. Our sample spans a wide range of galaxy properties: derived SFRs range from less than 10 to a few hundred solar masses per year; values of $M_\star$ range from 10$^9$ to 10$^{10}$ $M_\sun$ with a median of 5.6 $\times$ 10$^9$\,$M_\sun$. Comparing the specific star formation rate ($\phi$ $\equiv$ SFR/$M_\star$) of our sample as a function of $M_\star$ to other representative types of galaxies (distant red galaxies, Ly$\alpha$ emitters, Lyman break galaxies, submillimeter galaxies and $z$ $\sim$ 2 galaxies from the Great Observatories Origins Deep Survey-North field), we find that gamma ray burst hosts are among those with the highest $\phi$.

\end{abstract}


\section{Introduction}
\label{intro}

A canonical model is well established for long duration gamma ray bursts (GRB): association with stellar core collapse events and hence with high mass star formation \citep*[e.g.][]{hjorth03,stanek03,zeh04,campana06}. The emerging picture, however, is complex. Most GRB host galaxies are faint and blue \citep{fruchter99,le floc'h03}. A few hosts show tentative evidence of very high star formation rates \citep*[SFRs;][]{chary02,berger03}, but their optical properties do not appear typical of the galaxies found in blind submillimeter galaxy surveys \citep*{tanvir04,christensen04,fruchter06}.

The {\em Spitzer} \citep{werner04} IRAC \citep[Infrared Array Camera;][]{fazio04} and MIPS \citep[Multiband Imager Photometer for {\em Spitzer};][]{rieke04} photometry, together with optical, near IR, submillimeter and radio observations, can help establish how GRB hosts relate to other high redshift galaxy populations. This is essential if we are to understand the full range of properties of star forming galaxies at high redshifts and exploit the potential of GRBs as more general probes of cosmic star formation.

In this Letter we study a subsample of the 16 GRB hosts observed with {\em Spitzer} by \citet{le floc'h06}. We compute SFRs, dust masses ($M_{\rm dust}$) and stellar masses ($M_\star$) for hosts at redshifts close to 1. This redshift is particularly relevant because it has been argued that the global SFR peaks there \citep{madau98}. To determine SFRs we fit full (optical to radio) spectral energy distributions (SED), which allows us to calculate the most accurate and robust values for the SFR in a sample of GRB host galaxies to date. At $z$ $\sim$ 1 $M_{\rm dust}$ is constrained by 850\,$\mu$m SCUBA observations. To determine $M_\star$ we fit observed 4.5\,$\mu$m fluxes. The $M_\star$ estimator is well calibrated for $z$ $\sim$ 1 \citep{labbé05} since the 4.5\,$\mu$m observed wavelength corresponds to the rest frame $K$ band. This gives us, for the first time, an accurate value of the stellar mass in those hosts. We assume an $\Omega_m$ = 0.3, $\Omega_\Lambda$ = 0.7 cosmology with $H_0$ = 70\,km s$^{-1}$ Mpc$^{-1}$.

\begin{deluxetable*}{ccccccccccc}
    \setlength{\tabcolsep}{0.03in}
\tabletypesize{\small}
   \tablewidth{0pt}
 \tablecaption{Hosts: flux densities, star formation rates, dust and stellar masses
        \label{flux:sfr:mass}
              }

\tablehead{
                        \colhead{}            & \multicolumn{2}{c}{Redshift}               & \colhead{}                                      & \colhead{}                                      & \colhead{}                                     & \multicolumn{3}{c}{SFR\tablenotemark{b}~~~($M_\sun$\,yr$^{-1}$)}                           & \colhead{}               & \colhead{}               \\
\cline{2-3} \cline{7-9} \colhead{}            & \colhead{}    & \colhead{}                 & \colhead{$f_{\rm 4.5\,\mu m}$\tablenotemark{a}} & \colhead{$f_{\rm 8.0\,\mu m}$\tablenotemark{a}} & \colhead{$f_{\rm 24\,\mu m}$\tablenotemark{a}} & \colhead{}                       & \colhead{}                 & \colhead{}                 & \colhead{$M_{\rm dust}$} & \colhead{$M_\star$}      \\
                        \colhead{GRB Host}    & \colhead{$z$} & \colhead{Ref.}             & \colhead{($\mu$Jy)}                             & \colhead{($\mu$Jy)}                             & \colhead{($\mu$Jy)}                            & \colhead{UV$_{\rm cont}$}        & \colhead{Ref.}             & \colhead{$L_{\rm 8-1000}$} & \colhead{(10$^7M_\sun$)} & \colhead{(10$^9M_\sun$)}
          }

\startdata
            \object[GRB970508]{970508} ...... & 0.83          & \citetalias{bloom98}       & $<$3.1                                          & $<$17                                           & $<$82                                          & \phn  2.5\tablenotemark{c}       & \citetalias{chary02}       & 0.4--26                    & 0.3--1.1                 & \phn 1.5                 \\
            \object[GRB970828]{970828} ...... & 0.96          & \citetalias{djorgovski01}  &    3.9 $\pm$ 1.1                                & $<$18                                           &    94 $\pm$ 17                                 & \phn  1.1\tablenotemark{d}       & \citetalias{djorgovski01}  &       30 $\pm$ 8           &       1.3                & \phn 2.5                 \\
            \object[GRB980613]{980613} ...... & 1.10          & \citetalias{djorgovski03}  &     37 $\pm$ 1                                  &    33 $\pm$ 8                                   &   169 $\pm$ 36                                 &      70\tablenotemark{c}\phd\phn & \citetalias{chary02}       &      428 $\pm$ 51          &      19                  &     31\phd\phn           \\
            \object[GRB980703]{980703} ...... & 0.97          & \citetalias{djorgovski98}  &     11 $\pm$ 2                                  & $<$24                                           & $<$85                                          &      30\tablenotemark{c}\phd\phn & \citetalias{chary02}       & 3.8--226                   & 2.7--10                  & \phn 7.2                 \\
            \object[GRB981226]{981226} ...... & 1.11          & \citetalias{christensen05} &    4.5 $\pm$ 1.4                                & $<$31                                           & $<$87                                          & \phn  1.2\tablenotemark{d}       & \citetalias{christensen05} & 1.0--84                    & 0.7--3.7                 & \phn 3.9                 \\
            \object[GRB990705]{990705} ...... & 0.84          & \citetalias{le floc'h02}   &     19 $\pm$ 1                                  & $<$18                                           &   159 $\pm$ 31                                 & $\sim$5\tablenotemark{d}\phn\phn & \citetalias{le floc'h02}   & 4.5--173                   & 3.2--7.7                 & \phn 9.2                 \\
  \enddata

\tablerefs{
           \citepalias{bloom98}        \citealt{bloom98};
           \citepalias{djorgovski01}   \citealt{djorgovski01};
           \citepalias{djorgovski03}  \citealt*{djorgovski03};
           \citepalias{djorgovski98}   \citealt{djorgovski98};
           \citepalias{christensen05}  \citealt{christensen05};
           \citepalias{le floc'h02}    \citealt{le floc'h02};
           \citepalias{chary02}        \citealt{chary02}.
          }

\tablenotetext{a}   {Upper limits are quoted at the 3$\sigma$ level, while errors are 1$\sigma$.}
\tablenotetext{b}   {We have corrected all the SFRs quoted to account for differences in their IMF scales with respect to our choice of a Salpeter IMF \citep[0.1--100 $M_\sun$;][]{salpeter55}.}
\tablenotetext{c}   {Corrected using the $\beta$ slope technique \citep[][and references therein]{chary02}, typically larger than the Balmer lines decrement correction.}
\tablenotetext{d}   {Uncorrected for internal extinction; \citealt{christensen05} argue for no extinction in the host of \object[GRB981226]{GRB 981226}.}

\end{deluxetable*}

\section{Data}
\label{data}

From the sample of 16 GRB hosts (GTO programme 76)\ we selected six that have $z$ $\sim$ 1 (1$ ^{+0.11}_{-0.17}$). The reader is referred to \citet{le floc'h06} for a detailed description of the full data set. Each host has been imaged with IRAC and MIPS. IRAC observations were 4.5\,$\mu$m (300\,s per host; scale = 1.220\arcsec\,pixel$^{-1}$; FoV = 5.21 $\times$ 5.21\,arcmin$^2$; instrumental PSF FWHM = 1.98\arcsec) and 8.0\,$\mu$m (300\,s per host; scale = 1.213\arcsec\,pixel$^{-1}$; FoV = 5.18 $\times$ 5.18\,arcmin$^2$; instrumental PSF FWHM = 1.72\arcsec). MIPS observations were 24\,$\mu$m (420\,s per host; scale = 2.45\arcsec\,pixel$^{-1}$; FoV = 5.23 $\times$ 5.23\,arcmin$^2$; instrumental PSF FWHM $\sim$ 6\arcsec).

We used official {\em Spitzer} Post Basic Calibrated Data products (carefully verified with our own reductions). Host extraction was based on archival imagery world coordinate system calibration and visually confirmed with optical comparison images from the literature. The median separation between the host centroid in each $Spitzer$ image and the best set of coordinates published was about 1\arcsec. We measured the flux densities over a circled area of radius 2\,pixels in IRAC and 3\,pixels in MIPS. Aperture corrections were then applied to account for the extended size of the PSF. Our photometry, presented in Table \ref{flux:sfr:mass}, is consistent with \citet{le floc'h06}.

\begin{figure}
      \epsscale{1.25}
               \plotone{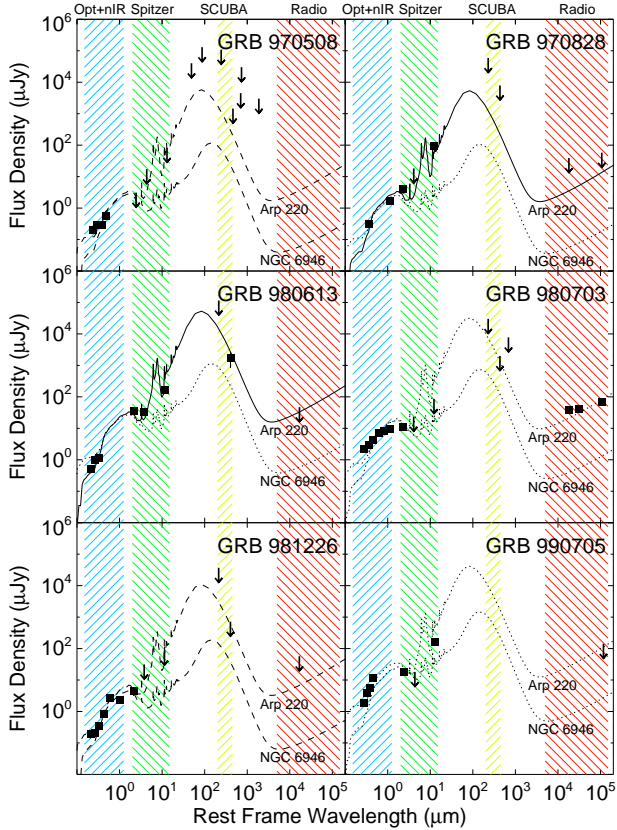}
                       \caption{SED fits of the six GRB host galaxies. {\em Filled squares}: detections ({\em error bars within the squares}). {\em Arrows}: 3 $\sigma$ upper limits (value marked by the base of the arrow). For each host in our sample we plotted the fit to the template that yielded the highest (\protect\object{Arp 220}) and the lowest (\protect\object{NGC 6949}) SFR value. {\em Solid lines}: best fit. {\em Dashed lines}: fits we could not discriminate between. {\em Dotted lines}: fits inconsistent with the data, shown for illustrative purposes. {\em Datapoints}: optical/near IR from \citet{vreeswijk99}, \citet{sokolov01}, \citet{djorgovski01}, \citet{le floc'h02} and \citet{christensen05}. Thermal IR from our photometry of {\em Spitzer} archival data. Far IR from \citet{hanlon00}. Submillimeter from \citet{smith99}, \citet{tanvir04} and our photometry of SCUBA archival data. Radio from \citet{bremer98}, \citet{shepherd98}, \citet*{berger01}, \citet{djorgovski01} and \citet{berger03}.
                       \label{arp220.ngc6946}
                      }
  \end{figure}

\section{Spectral Energy Distribution Fits}
\label{sed fits}

Photometry available for our sample includes our {\em Spitzer} flux densities and data points from the literature, spanning from optical to radio wavelengths. We scaled a set of SED templates to fit these data: \object{Arp 220}, an archetypal ultraluminous IR galaxy (ULIRG); \object{NGC 6946}, a well characterised, blue, star forming galaxy \citep[both][]{silva98}; and 64 SED templates ranging from starbursts to quiescent ones \citep{dale01,dale02}. The spectral width of the different bandpass filters was taken into account. Fitting was evaluated by means of a weighted least squares method.

Figure \ref{arp220.ngc6946} shows our models. The \object{Arp 220} template best fits the hosts of GRBs \object[GRB970828]{970828} and \object[GRB980613]{980613}. For these two hosts the data rule out the other SED templates; we have plotted the \object{NGC 6946} template for comparison. For the hosts of GRBs \object[GRB970508]{970508} and \object[GRB981226]{981226} we could not unambiguously discriminate a best fitting SED template, so we plotted those models yielding the highest and lowest values for the SFR (see $\S$ \ref{sfr} and Table \ref{flux:sfr:mass}). Model degeneracy was expected in these two cases because we only have optical--IR data. For the remaining two hosts we also plotted those models yielding the highest and lowest values for the SFR. For \object[GRB980703]{GRB 980703}, the \object{Arp 220} template approximately reproduced the radio flux densities but was inconsistent with the 24\,$\mu$m upper limit, while the \object{NGC 6946} template was consistent with the 24\,$\mu$m upper limit but underestimated the radio flux densities. For \object[GRB990705]{GRB 990705}, the \object{Arp 220} template overestimated the 24\,$\mu$m flux density and was marginally inconsistent with the 8\,$\mu$m upper limit, while the \object{NGC 6946} template underestimated the 24\,$\mu$m flux density. Reproducing the SEDs of these two galaxies is problematic and may require dust with properties different from those in our templates. The \citet{dale01} and \citet{dale02} templates yielded SFR values between those of the \object{Arp 220} and \object{NGC 6946} models for all the hosts in our sample.

For each host we may be fitting subcomponents that differ in their properties \citep{charmandaris04}. For instance, for \object[GRB980613]{GRB 980613} some components detected in the optical/near IR bands \citep*{hjorth02,djorgovski03} are offset by $>$2.5\arcsec\ from the $Spitzer$ centroid \citep*{le floc'h06}. While such effects might induce some scatter in the SEDs we can still utilise the SED templates as powerful diagnostic tools for the SFRs of the sample.

\subsection{Star Formation Rates}
\label{sfr}

Using our SED fitting models, we calculated the SFR for each host using IR luminosities. We converted flux densities into luminosity densities using $L_\nu(\nu_{\rm rest}) = 4{\pi}D_L^2f_\nu(\nu_{\rm observed})/(1+z)$ \citep{hogg02}, where $D_L$ is the luminosity distance, and scaled the SED templates to match the data points. IR luminosities ($L_{\rm 8-1000}$) were obtained integrating under the scaled SED templates from 8 to 1000\,$\mu$m (rest frame). This wavelength range was chosen so the SFRs could be computed using SFR$(M_\sun\,{\rm yr}^{-1}) = 4.5 \times 10^{-44}L_{\rm IR}$\,erg\,s$^{-1}$ \citep{kennicutt98}. The results are summarised in Table \ref{flux:sfr:mass}. Errors quoted are statistical and assume the template is a good representation of the data. In addition there may be significant systematic errors related, for example, to how well the template represents the actual SED \citep{micha?owski06}, the $L_{\rm 8-1000}$ to SFR conversion \citep{kennicutt98} (both of the same order, $\sim$30\%), or a factor of two from the choice of initial mass function \citep[IMF;][]{erb06a}.

For the hosts of GRBs \object[GRB970508]{970508}, \object[GRB980703]{980703}, \object[GRB981226]{981226} and \object[GRB990705]{990705}, the lower end of their SFR ranges indicates low star formation, consistent with the estimates from the UV continuum and, in \object[GRB981226]{GRB 981226}, with no internal extinction \citep*{christensen05}. The host of \object[GRB970828]{GRB 970828} is a moderately star forming galaxy, in good agreement with \citet{le floc'h06}. The host of \object[GRB980613]{GRB 980613} is characterised by high star formation activity. Our SFR value for this host is $\sim$5 times higher than the one obtained by \citet{le floc'h06} with a lower uncertainty, because we have fitted the entire SED, as opposed to only the flux density at 24\,$\mu$m. We verified that if we base our calculations exclusively on the 24\,$\mu$m flux densities we reproduce the results in \citet{le floc'h06} for all hosts.

\subsection{Dust Masses}
\label{dust}

Dust emission dominates submillimeter wavelengths. The total $M_{\rm dust}$ in a galaxy can be estimated from its rest frame 450\,$\mu$m flux density: $M_{\rm dust} = S_{\nu}D_L^2/(1+z)\kappa(\nu)B(\nu,T)$; where $S_\nu$ is the flux density at an observed wavelength corresponding to the rest frame wavelength of 450\,$\mu$m at $z$ = 1, interpolated from the fitted SED templates; $\nu$ is the frequency (666.21\,GHz) corresponding to a wavelength of 450\,$\mu$m; $\kappa$ $\propto$ $\nu^\beta$ is the mass absorption coefficient with $\beta$ being the dust emissivity index; and B($\nu,T$) is the Planck function \citep{taylor05}. We assumed optically thin dust emitting a grey spectrum. This method yields statistical errors of $\sim$25\% \citep{taylor05}. The ranges in $M_{\rm dust}$ are estimates of the systematic, model dependent error. Our results are listed in Table \ref{flux:sfr:mass}. The derived median, $M_{\rm dust}$ = 8 $\times$ 10$^6$\,$M_\sun$ (calculated from the lowest $M_{\rm dust}$ value for each host), is consistent with the distribution for starburst galaxies (4 $\times$ 10$^5$ to 7 $\times$ 10$^8$ $M_\sun$) found by \citet{taylor05}.

\section{Stellar Masses}
\label{m*}

We estimated $M_\star$ from rest frame $K$ band fluxes \citep[e.g.][]{glazebrook04}. $M/L_K$ depends to some extent on the composition of the stellar population \citep*{portinari04} or, according to \citet{labbé05} who used \citet{bruzual03} with a Salpeter IMF, on the rest frame $U-V$ colour, age and $M_\star$. GRB hosts are blue, young and faint \citep[e.g.][]{le floc'h03,berger03,christensen04}; thus, we assume $M/L_K$ $\sim$ 0.1 (lowest detection in \citealt{labbé05} is 0.16), obtaining a robust lower limit. Table \ref{flux:sfr:mass} summarises our $M_\star$ estimates. \citet{van der wel06} examined redshift dependent systematics in determining $M_\star$ from broad band SEDs. They found no significant bias for \citet{bruzual03} models with a Salpeter IMF \citep{salpeter55}.

\begin{figure}
      \epsscale{1.16}
               \plotone{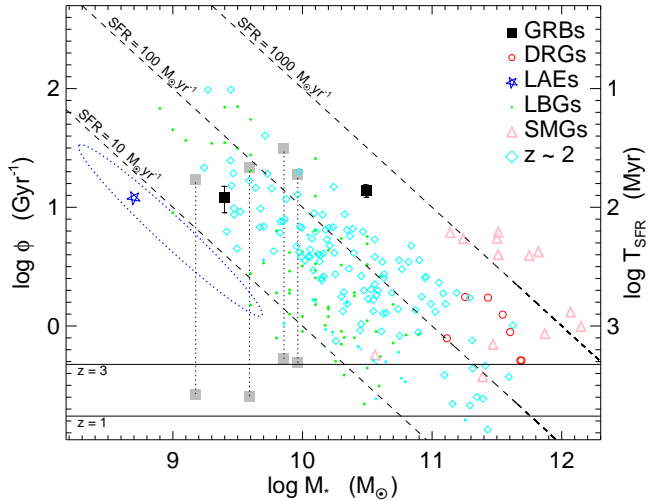}
                       \caption{$\phi$ as a function of $M_\star$ for our GRB host galaxy sample and other representative types of galaxies \citep[for a similar plot see][]{erb06b}. {\em Black squares}: SFR values constrained with a best fit model. {\em Grey squares}: highest/lowest SFR values for those hosts for which a best fit model could not be established. {\em Dashed diagonals}: SFRs of 1000, 100 and 10 $M_\sun$\,yr$^{-1}$ respectively. {\em Right axis}: SFR timescale ($T_{\rm SFR}$ = $M_\star$/SFR), the inverse of $\phi$. On this scale the solid horizontals represent the age of the universe for the marked redshift. Our six hosts clearly have $T_{\rm SFR}$ $<$ t$_{\rm universe}$, allowing for a history of constant star formation. The distribution of our sample in parameter space suggests that GRBs trace galaxies that are not selected with other techniques. {\em Data points}: GRBs from this Letter. DRGs from \citet{van dokkum04}. LAEs from \citet{gawiser06}. The point plotted represents an average value of the LAE population as a whole, obtained from stacked photometry , and the dotted ellipse its uncertainty. Spectroscopically confirmed LBGs from \citet{shapley01} and \citet{barmby04}. SMGs from \citet[][$M_\star$]{chapman05} and \citet[][SFR]{borys05}, where we have considered $L_{\rm BOL}$ $\simeq$ $L_{\rm far IR}$, then applied the \citet{kennicutt98} calibration. $z$ $\sim$ 2 from \citet{erb03} and \citet{reddy06}.
		         \label{galaxy.samples}
                      }
  \end{figure}

\section{Discussion}
\label{discussion}

We have found the hosts in our sample to span a wide range of properties. Their SEDs are fitted with templates that vary from a blue, star forming galaxy to a ULIRG. Their SFRs and $M_\star$ are quite different, ranging from the host of \object[GRB980613]{GRB 980613}, which is forming a few hundred solar masses a year with $M_\star$ = 3 $\times$ 10$^{10}$\,$M_\sun$, to the host of \object[GRB970508]{GRB 970508}, which is forming of the order of 10\,$M_\sun$ a year with $M_\star$ = 1.5 $\times$ 10$^9$\,$M_\sun$.

We find that our SFR values are significantly higher (up to a factor of 30) than the lower limits from the rest frame UV continuum emission ($L_{\rm UV}$; see Table \ref{flux:sfr:mass}) and higher (by a factor of 6) than those corrected with the $\beta$ slope technique \citep{chary02}. GRBs \object[GRB970828]{970828}, \object[GRB980613]{980613} and \object[GRB980703]{980703} illustrate how even the best estimates of dust extinction in a galaxy from the UV slope may fall short, not only for ULIRGs \citep{chary01}, but also for LIRGs (i.e. 10$^{12}$\,$L_\sun$ $>$ $L_{\rm 8-1000}$ $\geq$ 10$^{11}$\,$L_\sun$). Our SFR value for \object[GRB981226]{GRB 981226} is consistent with no internal extinction \citep{christensen05}.

The specific star formation rate ($\phi$ $\equiv$ SFR/$M_\star$) gives an indication of how intensely star forming a galaxy is. In Figure \ref{galaxy.samples} we have plotted $\phi$ versus $M_\star$ for ours and five other representative galaxy samples: distant red galaxies (DRG), Ly$\alpha$ emitters (LAE), Lyman break galaxies (LBG), submillimitre galaxies (SMG) and an ensemble of optically selected, $z$ $\sim$ 2 galaxies from the Great Observatories Origins Deep survey-North field. GRB hosts have some of the highest $\phi$ values, as previously suggested by \citet{christensen04}. $\phi^{-1}$ represents the SFR timescale, so high $\phi$ values are indicative of GRBs tracing young, starbursting galaxies.

The different methods that have been used to derive the SFRs of the various samples plotted in Figure \ref{galaxy.samples} likely introduce systematic offsets. Likewise, to determine $M_\star$, model fitting to the full SED could better account for variations in $M/L_K$. For high SFR galaxies, ongoing star formation contributes to the near IR emission, especially as they lack an old stellar population. High $\phi$ galaxies are particularly vulnerable to this effect. Nevertheless, Figure \ref{galaxy.samples} illustrates where, within the larger picture, our hosts fall.

We have shown the capabilities of IR observations to characterise GRB host galaxies and have compared values of $\phi$ for different types of galaxies irrespective of their redshift. Unfortunately, the selection effects are difficult to quantify for the present small sample, which therefore does not allow a robust statistical analysis. The next step to increase our understanding of GRB hosts will be to extend our mid and far IR observed sample, as a larger, well selected one will tell us more about the span of host properties (i.e. SFR, $M_{\rm dust}$ and $M_\star$). Future work should include full population synthesis modelling and address the redshift dependence of $\phi$.

\acknowledgments

We thank J. Baradziej, B. Cavanagh, F. Economou, \'A. El\'{\i}asd\'ottir, E. Gawiser, T.R. Greve, T. Jenness, K. Nilsson, R. Priddey, E. Ramirez-Ruiz and S. Toft for insightful comments, and our anonymous referee for genuinely helping us to improve the manuscript. The Dark Cosmology Centre is funded by the Danish National Research Foundation. JMCC gratefully acknowledges support from the Instrumentcenter for Dansk Astrofysik and the Niels Bohr Instituttet's International PhD School of Excellence. JG was funded in part by Spain's AyA 2.004-01515 and ESP 2.002-04124-C03-01 grants. The authors acknowledge benefits from collaboration within EU's FP5 RTN ``GRBs: an enigma and a tool" (HPRN-CT-2002-00294). This research has made use of NASA's Astrophysics Data System.

{\it Facilities:} \facility{Spitzer (IRAC, MIPS)}.

\end{document}